\documentclass[aps,prb,twocolumn,amsmath,amssymb,superscriptaddress,citeautoscript,longbibliography,floatfix]{revtex4-2}%

\usepackage{graphicx}
\usepackage{dcolumn}
\usepackage{bm}
\usepackage[english]{babel}
\usepackage{color}
\usepackage{multirow}
\usepackage{natbib}
\usepackage{hyperref}

\begin{document}

\title{High sensitivity heat capacity measurements on Sr$_2$RuO$_4$ under uniaxial pressure}

\author{Y.-S.\ Li}
\affiliation{Max Planck Institute for Chemical Physics of Solids,  N\"{o}thnitzer Str.\ 40, 01187 Dresden, Germany}
\affiliation{Scottish Universities Physics Alliance, School of Physics and Astronomy, University of St Andrews, St Andrews, UK}

\author{N.\ Kikugawa}
\affiliation{National Institute for Materials Science, Tsukuba 305-0003, Japan}

\author{D.\ A.\ Sokolov }
\affiliation{Max Planck Institute for Chemical Physics of Solids,  N\"{o}thnitzer Str.\ 40, 01187 Dresden, Germany}

\author{F.\ Jerzembeck }
\affiliation{Max Planck Institute for Chemical Physics of Solids,  N\"{o}thnitzer Str.\ 40, 01187 Dresden, Germany}

\author{A.\ S.\ Gibbs  }
\affiliation{ISIS facility, STFC Rutherford Appleton Laboratory, Chilton, Didcot, OX11 0QX, UK}

\author{Y.\ Maeno  }
\affiliation{Department of Physics, Graduate School of Science, Kyoto University, Kyoto, Japan}

\author{C.\ W.\ Hicks }
\email{Clifford.Hicks@cpfs.mpg.de}
\affiliation{Max Planck Institute for Chemical Physics of Solids,  N\"{o}thnitzer Str.\ 40, 01187 Dresden, Germany}

\author{M.\ Nicklas}
\email{Michael.Nicklas@cpfs.mpg.de}
\affiliation{Max Planck Institute for Chemical Physics of Solids,  N\"{o}thnitzer Str.\ 40, 01187 Dresden, Germany}

\author{A.\ P.\ Mackenzie }
\email{Andy.Mackenzie@cpfs.mpg.de}
\affiliation{Max Planck Institute for Chemical Physics of Solids,  N\"{o}thnitzer Str.\ 40, 01187 Dresden, Germany}
\affiliation{Scottish Universities Physics Alliance, School of Physics and Astronomy, University of St Andrews, St Andrews, UK}

\date{\today}

\begin{abstract}
A key question regarding the unconventional superconductivity of Sr$_2$RuO$_4$ remains whether the order parameter is single- or two-component.  Under a hypothesis of two-component superconductivity, uniaxial pressure is expected to lift their degeneracy, resulting in a split transition.  The most direct and fundamental probe of a split transition is heat capacity.  Here, we report development of new high-frequency methodology for measurement of heat capacity of samples subject to large and highly homogeneous uniaxial pressure.  We place an upper limit on the heat capacity signature of any second transition of a few per cent of the primary superconducting transition. The normalized jump in heat capacity, $\Delta C/C$, grows smoothly as a function of uniaxial pressure, but we find no qualitative evidence of a pressure-induced order parameter transition.
Thanks to the high precision of our measurements, these findings place stringent constraints on theories of the superconductivity of Sr$_2$RuO$_4$.

\end{abstract}

\maketitle

\section{INTRODUCTION}

The layered ruthenate superconductor Sr$_2$RuO$_4$ is a material of considerable importance to the fields of correlated electron physics and unconventional superconductivity.  When the superconductivity of Sr$_2$RuO$_4$ was discovered in 1994 \cite{Maeno1994}, it generated immediate interest because its crystal structure is the same as that of the parent compound of the cuprate superconductors, La$_2$CuO$_4$.  It was quickly realized that, unlike the superconducting cuprates, Sr$_2$RuO$_4$ is naturally stoichiometric and highly stable chemically \cite{Mackenzie2003}.  The superconductivity was seen to be extremely sensitive to non-magnetic disorder \cite{Mackenzie1998}, and careful purification led to crystals with mean free paths of over $1~{\rm \mu m}$, in which it was possible to study unconventional superconductivity in the clean limit and determine the parameters of the Fermi liquid metallic state with extremely high precision \cite{Maeno1994,Mackenzie1996,Maeno1997,Bergemann2000,Bergemann2003}.

A quarter of a century after the discovery of the superconductivity, the normal state physics of Sr$_2$RuO$_4$ is known to involve a rich and subtle interplay of Coulomb repulsion, spin orbit coupling and Hund's rule physics that is unique to $4d$ systems in which all these energy scales are similar \cite{Nomura2002,Haverkort2008,Raghu2010,Mravlje2011,Veenstra2014,Scaffidi2014,Tamai2019}.  In spite of intensive investigation, however, the superconductivity is less well understood.  There is no consensus on the mechanism of the pairing and even on the symmetry of the superconducting state.  Interpretation of experiments studying the superconductivity has often been referenced to a beautiful theoretical proposal that Sr$_2$RuO$_4$ might host an odd parity superconducting state, making it a quasi-two-dimensional analog of superfluid $^3$He \cite{Rice1995}.  For many years this was lent strong support by the observation that the nuclear magnetic resonance (NMR) Knight shift remained constant below the superconducting transition temperature ($T_c$) \cite{Ishida1998}.  A temperature-independent Knight shift is impossible in a clean superconductor with an even parity order parameter, and it also places strong constraints on the sub-class of odd parity states that can exist.  Among the simplest $p$-wave states, it favors one with a vector order parameter of the form $\mathbf{d}=\mathbf{\hat{z}}(k_x \pm ik_y)$, in which time reversal symmetry is broken at $T_c$.  Independent observations supported the existence of such a state \cite{Luke1998,Duffy2000,Kealey2000,Xia2006,Kidwingira2006}, and although other experiments gave worrying inconsistencies \cite{Kirtley2007,Hicks2010,Hassinger2017}, the $\mathbf{d}=\mathbf{\hat{z}}(k_x \pm ik_y)$ order parameter remained the most favored one for nearly twenty years \cite{Mackenzie2003,Maeno2012,Kallin2012}.

In an effort to better understand the overall phase diagram of Sr$_2$RuO$_4$, experimental attention has been paid in recent years to strain tuning of its properties.  This can be done using epitaxial strain imposed on thin films grown on substrates with different lattice parameters \cite{Burganov2016,Hsu2016}, and has also been achieved using novel techniques for applying uniaxial pressure that were first developed explicitly for the study of Sr$_2$RuO$_4$ superconductivity, and have subsequently been applied to a range of other materials \cite{Hicks2014a,Hicks2014b,Steppke2017,Barber2018,Watson2018,Barber2019,Stern2017,Kissikov2017,Luo2019,Kim2018}.  In Sr$_2$RuO$_4$, uniaxial pressure applied parallel to the crystalline $a$ axis makes a substantial change to the superconductivity, raising $T_c$ by a factor 2.3 from 1.5 to 3.5~K, and increasing the critical fields by a factor of 20 to 1.5~T for magnetic field applied parallel to the crystalline $c$ axis, and from 1.5 to 4.5~T for magnetic field applied in the $ab$ plane \cite{Steppke2017}.  These developments motivated a new study of the Knight shift as a function of uniaxial pressure, with surprising results \cite{Pustogow2019}.  Sample heating effects were proven to have affected the previous NMR experiments, and it is now fully accepted that the Knight shift in fact changes considerably on entry to the superconducting state \cite{Pustogow2019,Ishida2019}. A result from magnetic neutron scattering \cite{Duffy2000} that had appeared to reinforce the previous NMR experiment has also very recently been revised \cite{Petsch2020}. These new observations are completely inconsistent with the previously favored $\mathbf{d}=\mathbf{\hat{z}}(k_x \pm ik_y)$) order parameter.

\begin{figure}[b!]
\includegraphics[width=0.95\linewidth]{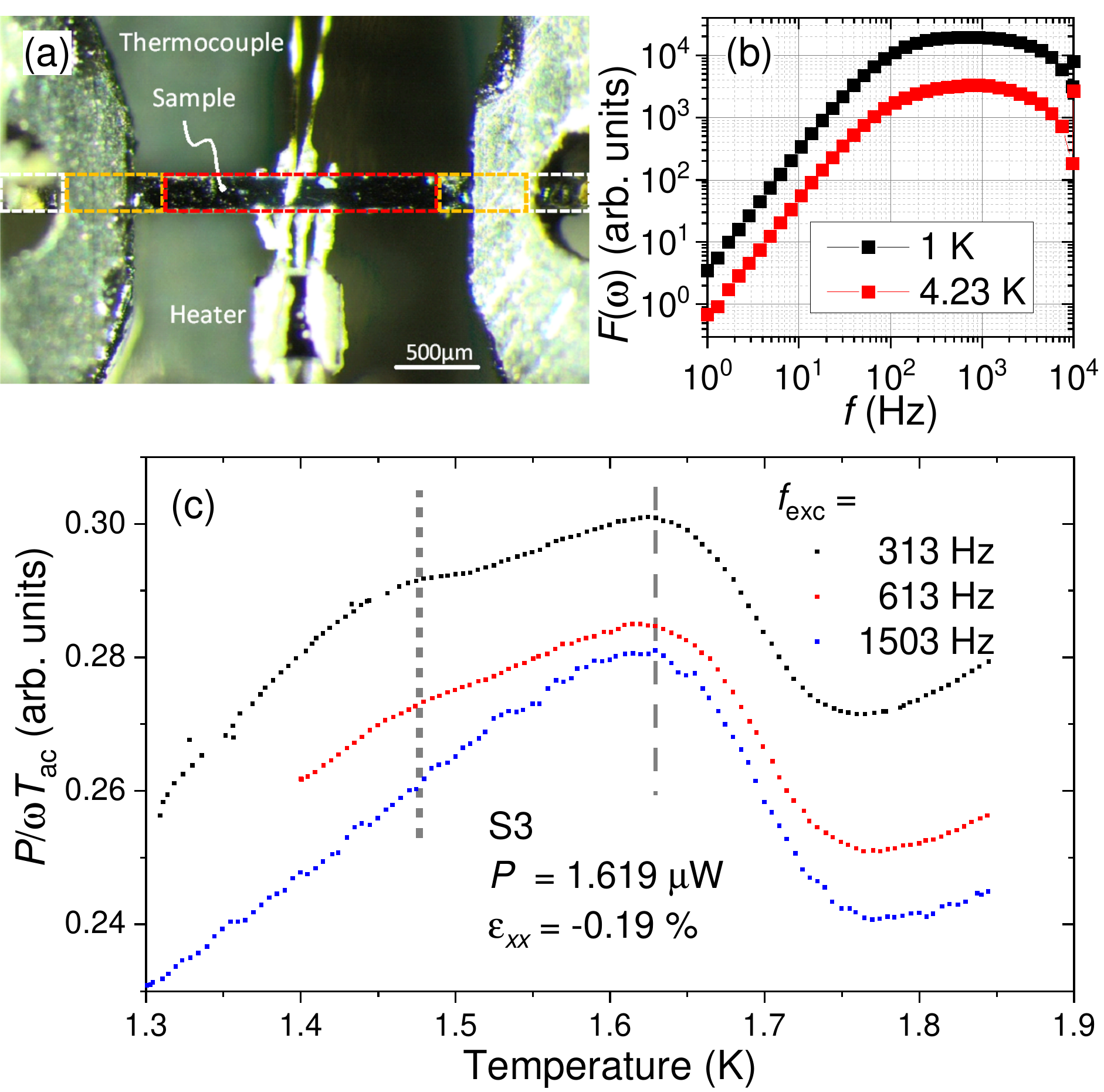}
\centering
\caption{
(a) A photograph of the setup of the heat capacity measurements under strain including heater and thermometer. The sample is glued between the jaws of the uniaxial pressure device. It allows the application of compressive and tensile strains. The red, yellow, and white dashed lines respectively indicate schematically the homogeneously strained, intermediate, and unstrained regions of the sample. (b) Frequency sweeps for S3 at $T = 1$~K and 4.23~K.  (c) Heat capacity measurements under a strain $-0.19\%$ for S3 at various excitation frequencies. The dashed (dotted) line indicates the superconducting transition from strained (unstrained) part of the sample.
}
\label{Config}
\end{figure}

The new measurements reinvigorate the quest to understand the superconductivity of Sr$_2$RuO$_4$.  They remove a worrying discrepancy between the critical field and Knight shift results \cite{Kittaka2014} that was discussed in some detail in Ref.\ \onlinecite{Mackenzie2017}, and rule out not just the $\mathbf{d}=\mathbf{\hat{z}}(k_x \pm ik_y)$ order parameter, but most other odd parity $p$-wave ones as well. This has in turn focused renewed attention on other key features of the superconducting state of Sr$_2$RuO$_4$. In particular, several new experiments have addressed again the question of whether the superconducting order parameter has two components, and separate pulse-echo and resonant ultrasound experiments both provide evidence that it does \cite{Benhabib2020,Ghosh2020}.  In addition, muon spin relaxation data \cite{Luke1998,Luke2000,Grinenko2020} consistently show spontaneous relaxation associated with superconductivity.  In combination with previous Kerr rotation measurements \cite{Xia2006}, the muon experiments provide evidence that the superconductivity has two components and breaks time reversal symmetry (TRS).

Most possible two-component order parameters and all plausible TRS-breaking ones for Sr$_2$RuO$_4$ are predicted to show superconducting transitions whose characteristic temperatures are split by the application of uniaxial pressure.  Experiments with a direct sensitivity to such a splitting are therefore of considerable topical importance. The new muon spin rotation experiment \cite{Grinenko2020} shows that the onset temperature of the relaxation splits off from the main superconducting transition when the crystal is pressured, an observation that invites comparison with results from thermodynamic probes.
Arguably the most fundamental of these is the heat capacity. As famously demonstrated in  UPt$_3$ \cite{Fisher1989}, the observation of a second heat capacity jump can reveal the presence of a second transition, and its magnitude provides information on the relationship between the two components, thereby placing strong constraints on theories of the order parameter symmetry.
A central challenge in advancing our understanding of the superconductivity of Sr$_2$RuO$_4$ is therefore to perform high-sensitivity measurements of its heat capacity under uniaxial pressure.  However, this is well known to present experimental difficulties; samples tend to break at relatively small uniaxial pressures, and even if this can be avoided, any measurement needs to deal with the generally large thermal conductance between the sample and its environment that results from the need to transmit large forces to the sample.

\section{EXPERIMENT}

To address the experimental challenge of studying the heat capacity under large uniaxial pressures, we employ a variation of known ac heat capacity measurement techniques \cite{Sullivan1968}. Heat capacity measurement has been combined with uniaxial pressure previously \cite{Jin1992,Reinders1994,Miclea2002,Zieve2004,Dix2009}, but with traditional, anvil-based uniaxial pressure cells.  Samples have been thermally isolated by using low thermal conductivity materials, such as stainless steel or superconducting NbTi, as piston or additional spacer.  However in previous anvil-based uniaxial stress measurements on Sr$_2$RuO$_4$, it did not prove practical to maintain high stress homogeneity \cite{Kittaka2010,Taniguchi2015}.  The superconducting transition broadened rapidly under applied pressure and the samples deformed plastically.  Either of these effects would severely hinder our goal of probing whether there is a splitting of the superconducting transition.  Therefore we apply force to the sample through a layer of epoxy \cite{Hicks2014a}, which acts as a conformal layer that dramatically improves stress homogeneity. However, it also makes heat capacity measurement more challenging, because the epoxy layer provides an unavoidable strong thermal link to the pressure cell.

In Fig.\ \ref{Config}a we show a photograph of a bar of single crystal Sr$_2$RuO$_4$ that has been carefully cut along its $[100]$ direction, polished, and then mounted within the jaws of the uniaxial pressure rig (see the Supplemental Material for additional information \cite{suppl}). Four independent experimental runs using different samples (S1 –- S4) proved the reliability of the results.

The governing relationship for measurements of the ac heat capacity $C_{ac}$ is
\begin{equation}
  C_{ac}=  \frac{P}{\omega T_{ac} } F(\omega).
\label{Cac}
\end{equation}
where $T_{ac}$ is the measured oscillation temperature amplitude, $P$ the power, and $\omega$ the angular frequency. $F(\omega)$ is a frequency response curve that characterizes the thermalization of the sample, and differs from sample to sample, because it depends on time constants, thermal conductances, and heat capacities of the system.  A measurement of $F(\omega)$ for one of our Sr$_2$RuO$_4$ samples is shown in Fig.\ \ref{Config}b, and is seen to follow the expected form for measurement under highly non-adiabatic conditions: $F(\omega)$ is reduced at low frequencies due to dissipation of temperature oscillations into the environment (here the body of the uniaxial pressure apparatus), and at high frequencies because the heater-sample-temperature sensor system does not thermalize. The plateau region between these limits is, in general, the region in which a heat capacity measurement can be made successfully.

For the current experiment, however, the conditions are still more demanding, because our goal is to measure the heat capacity of only a portion of the sample.  The nature of the apparatus means that only the central part of the Sr$_2$RuO$_4$ crystal is homogeneously strained. Force is transferred to the sample through the epoxy layer around the sample; as indicated by the colored lines in Fig.\ \ref{Config}a, the sample ends which are protruding beyond it are unstrained, and there are intermediate regions where the strain is built up. In the lower frequency part of the plateau in Fig.\ \ref{Config}b, temperature oscillations extend throughout the sample and all three regions are probed.
This is shown in Fig.\ \ref{Config}c, where the sample is placed under modest compression to raise $T_c$ of the central portion relative to the ends of the sample. At 313 and 613~Hz, in addition to the peak at $\approx 1.65$~K corresponding to the transition in the central portion, a smaller peak is visible at $\approx 1.45$~K, corresponding to the transition in the end portions. This feature shows that temperature oscillations extend into the sample ends.
To avoid this, one has to work at the high end of the feasible range of frequencies.
For this particular sample, a measurement frequency above $\sim1.5$~kHz was required to satisfy this more demanding criterion. The main results that we will present were taken at frequencies in the range $3.5 - 4$~kHz, and with low power input to minimize sample heating.  Working at those high frequencies gives a very low signal, with an r.m.s.\ thermocouple voltage of only $1 - 2$~nV, so low temperature passive amplification was employed to achieve an r.m.s. noise level of 20~pVHz$^{-1/2}$, ensuring a signal to noise ratio in excess of 50.

\section{RESULTS}

\begin{figure}[t!]
\includegraphics[width=0.95\linewidth]{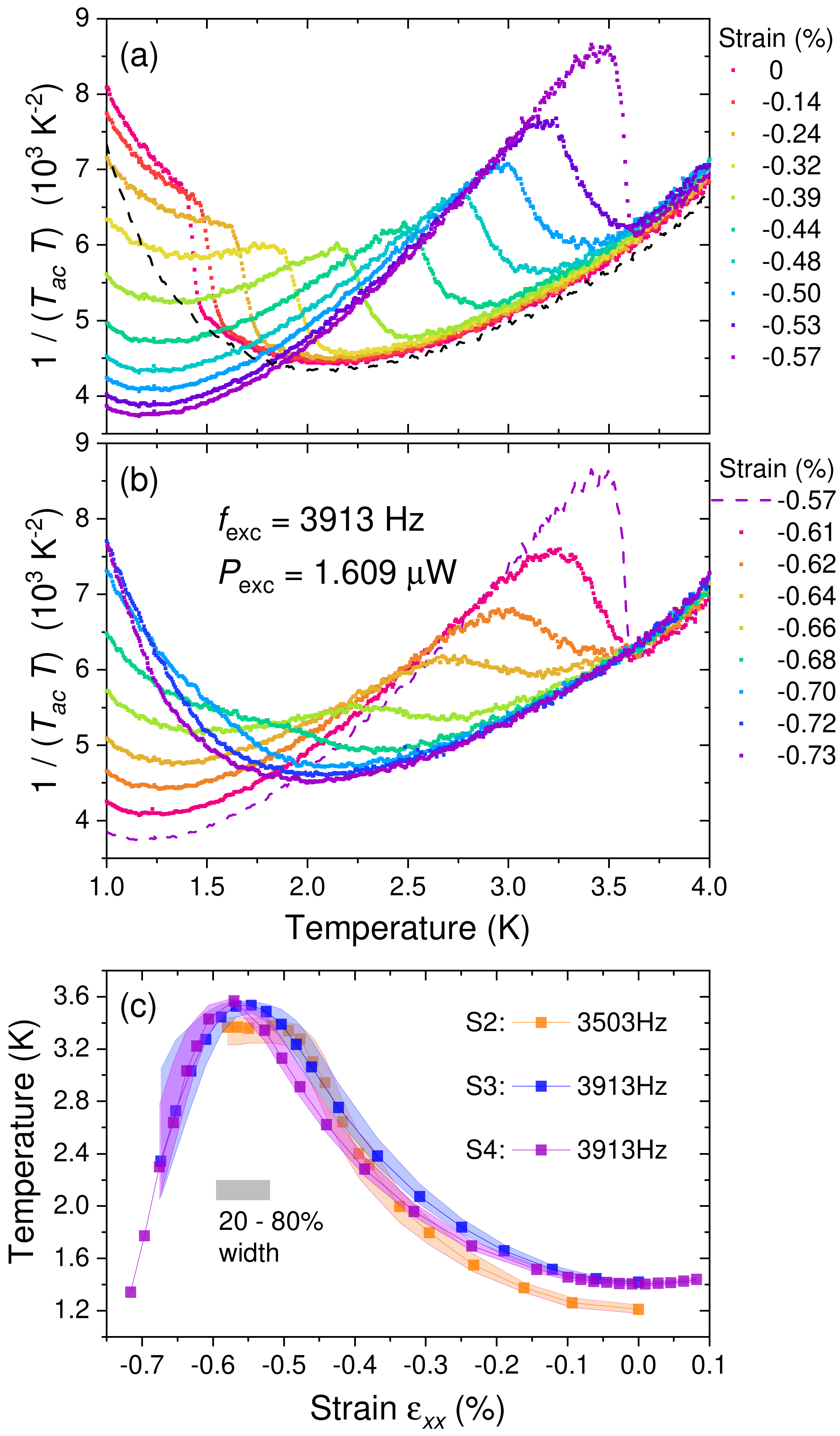}
\centering
\caption{
(a) Measurements at stresses before the peak in $T_c$. (b) Measurements at stresses beyond the peak in $T_c$. The dashed line in panel (a) is the heat capacity measurement at $\mu_0H_{\parallel c} = 0.1$~T and $\varepsilon_{xx}= 0$. The dashed line in panel (b) repeats the data at maximum $T_c$. (c) Superconducting transition temperature against strain for three experimental runs using different samples taken with the similar frequencies. $T_c$ of S4 has been probed at compressive and tensile strains. Solid points are $T_c$ taken as midpoints of the leading edge of the transitions ($50\%$ level). The colored area represents the width of the transitions from $20\%$ to $80\%$ levels.
}
\label{HC}
\end{figure}

A consequence of operating in the regime described above is that the probed volume is a function of both volume specific heat capacity $C_v$ and thermal conductivity $\kappa$, and so extracting a quantitatively accurate $C_v$ is not a trivial task. However the first physics questions that we aim to address here are qualitative: Is there a splitting of the superconducting transition as predicted for a TRS-breaking order parameter and is there a change in the jump height at $T_c$ at some uniaxial pressure that would indicate a stress-driven order parameter transition?

In Fig.\ \ref{HC} we show representative raw data demonstrating our capability of resolving the ac specific heat signal of the superconducting phase transition at strains below (Fig.\ \ref{HC}a) and above (Fig.\ \ref{HC}b) that at which $T_c$ is maximized.  The raw signal contains a background, which we illustrate for an unstrained sample by applying a small 0.1 T magnetic field to suppress the superconductivity (dotted line in Fig.\ \ref{HC}a). The breadth of the transition leading edge is due to residual strain inhomogeneity, due to bending and sample defects, and so is approximately proportional to the strain derivative of $T_c$ (Fig.\ \ref{HC}c). We estimate this residual inhomogeneity to be $8\%$ of the applied strain; see the Supplemental Material for additional information \cite{suppl}. Over most of the range of measurement the transition is well resolved, and the results for the strain dependence of the transition temperature (estimated as the mid-point of the leading edge of the heat capacity anomaly) are in excellent agreement with previous studies of the diamagnetic response (Fig.\ \ref{HC}c and Ref.\ \onlinecite{Steppke2017}).

\begin{figure}[t!]
\includegraphics[width=0.95\linewidth]{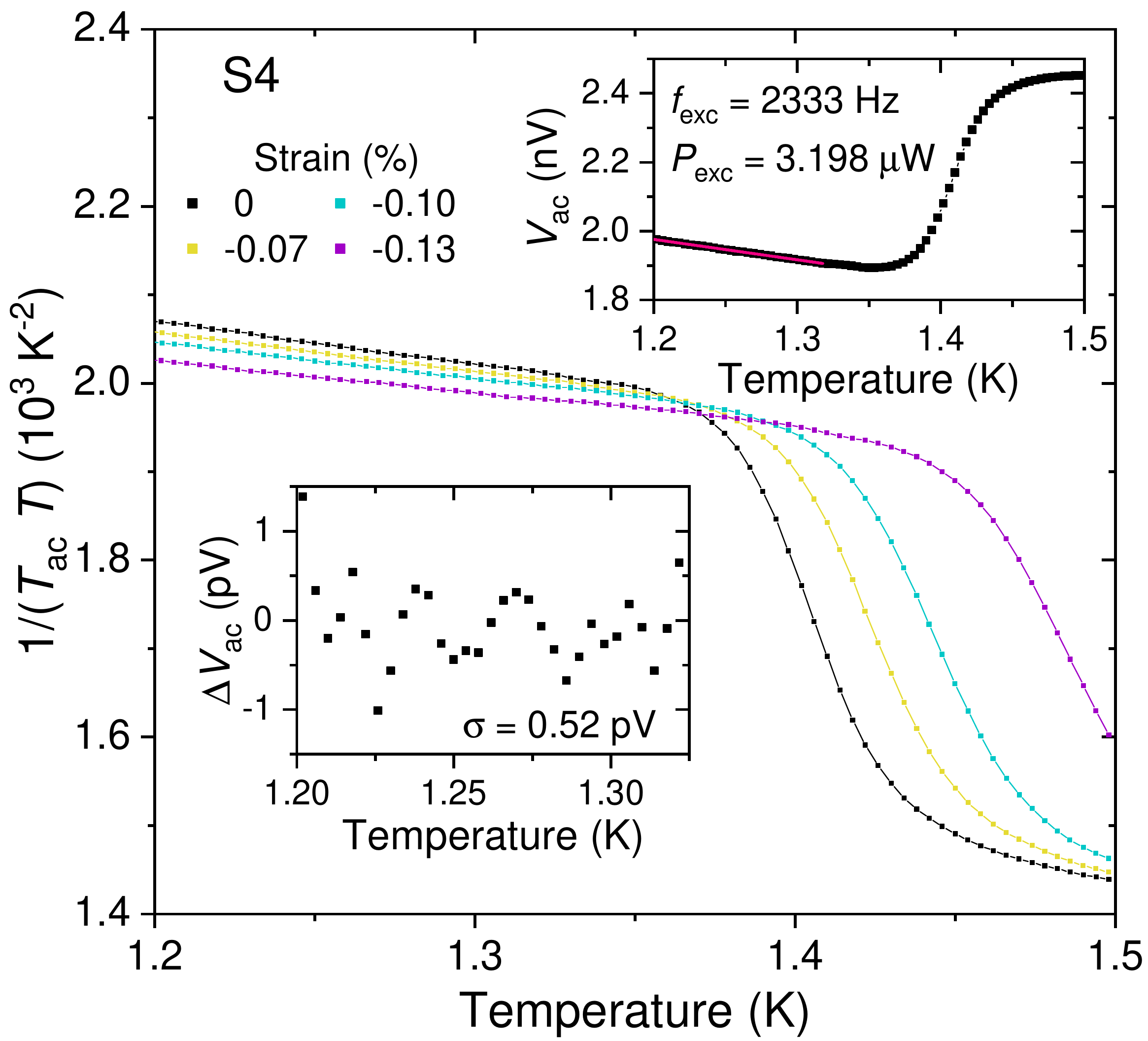}
\centering
\caption{
High resolution heat capacity measurements on sample S4. Measurements at several low strains. $P_{exc}$ was doubled and the measurements were repeated several times at each strain 20, 34, 18 and 20 times for the 0\%, -0.07\%, -0.1\% and -0.13\% curves, respectively. (upper inset) Averaged voltage readout from the thermocouple at zero strain. The red line is a linear fit to the data below $T_c$. (lower inset) The difference between the measured curve and the linear fit. The standard deviation is only about 0.52~pV.}
\label{HC_highRes}
\end{figure}

The data shown in Fig.\ \ref{HC} do not provide any indication of split transitions. Since one key motivation for our experiment was to look for transition splitting, we moved to higher excitation powers to maximize our signal-to-noise ratio. In Fig.\ \ref{HC_highRes} we present an analysis using a set of heat capacity recorded with a doubled excitation power. Additionally, the measurement runs were repeated and averaged up to 34 times providing higher quality heat capacity data with a much better signal-to-noise ratio. We restricted the measurements to the temperature range between 1.2 and 1.5 K since a potential second transition is expected at temperatures lower than the visible transition anomaly. There is also no signature of a second transition in the averaged curves. The noise of the averaged voltage at the thermocouple is only 0.52~pV (see lower inset of Fig.\ \ref{HC_highRes}). Following the analysis scheme described in the Supplemental Material \cite{suppl}, the detection limit relative to the primary transition is only 0.3\%. We note, however, that this value concerns the ability to resolve a sharp discontinuity.  If, as seems reasonable, a second transition had a similar width to the primary one, our detection limits would fall to a few per cent.  This still places very strong constraints on any theory proposing that strain splits the transition temperature of a two-component superconducting order parameter.

\begin{figure}[t!]
\includegraphics[width=0.9\linewidth]{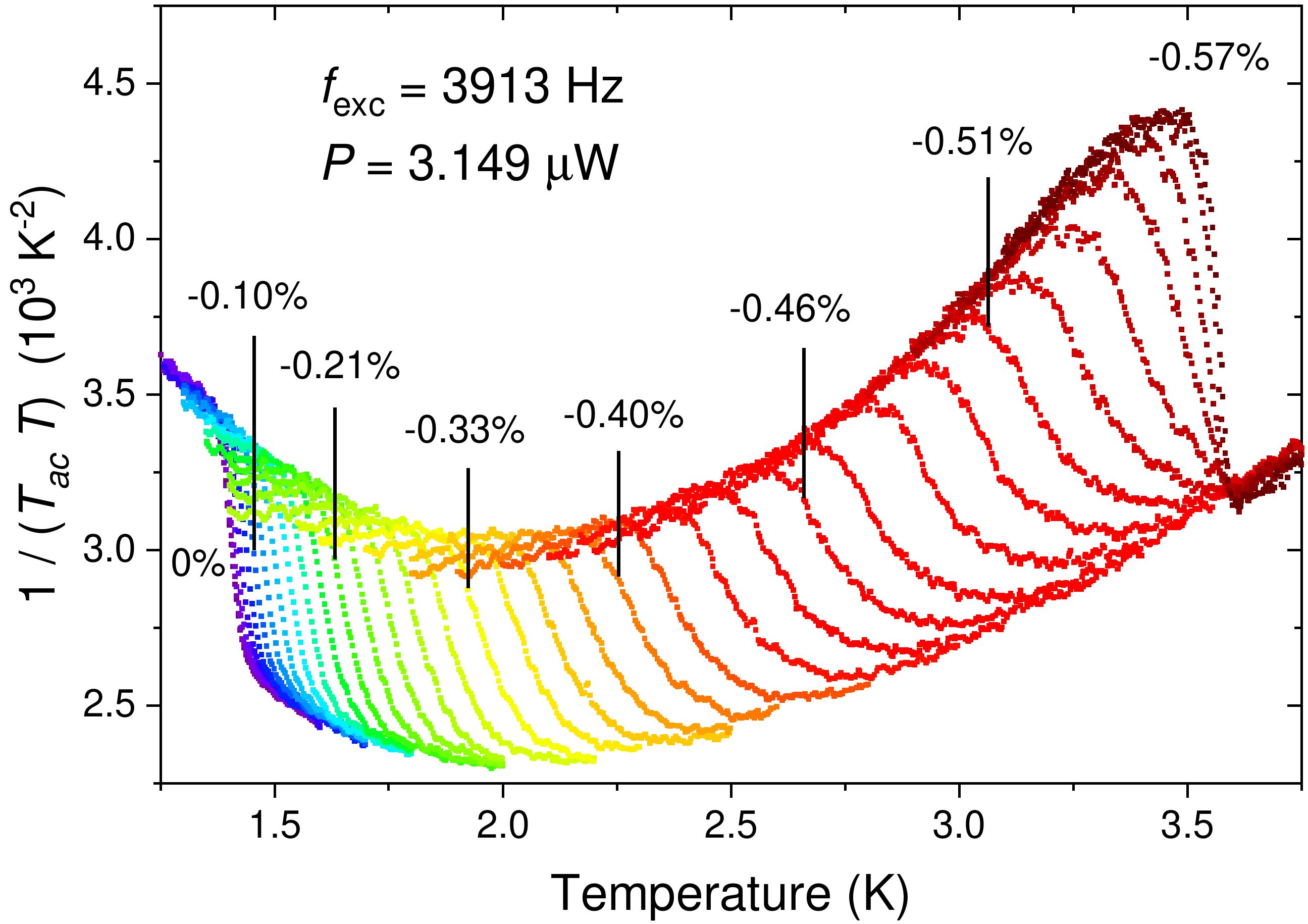}
\centering
\caption{
$(T_{ac}T)^{-1}$ against temperature for S4 at a series of small increments of compressive strains up to the peak in $T_c$. Smaller increments of strain and a larger power were used to reduce the noise and to allow more precise inspection of the presence of a transition between different order parameters inside the superconducting state.
}
\label{HC_smallsteps}
\end{figure}

Already the heat capacity data in Fig.\ \ref{HC} suggest a smooth evolution of the superconducting transition with stress. However, this is not the only possibility, because stress-tuned transitions between entirely different superconducting order parameters (even between those odd and even parity) can in principle occur, and might be signalled by a sudden jump in the size of the heat capacity anomaly at some critical stress.  To investigate this possibility more closely, we collected data at higher excitation powers and in fine stress steps. In Fig.\ \ref{HC_smallsteps} we present the result which, qualitatively, shows a smooth growth in the height of the heat capacity anomaly, but no sudden changes that would be indicative of a stress-induced transition between two different order parameters.

\begin{figure}[t!]
\includegraphics[width=0.9\linewidth]{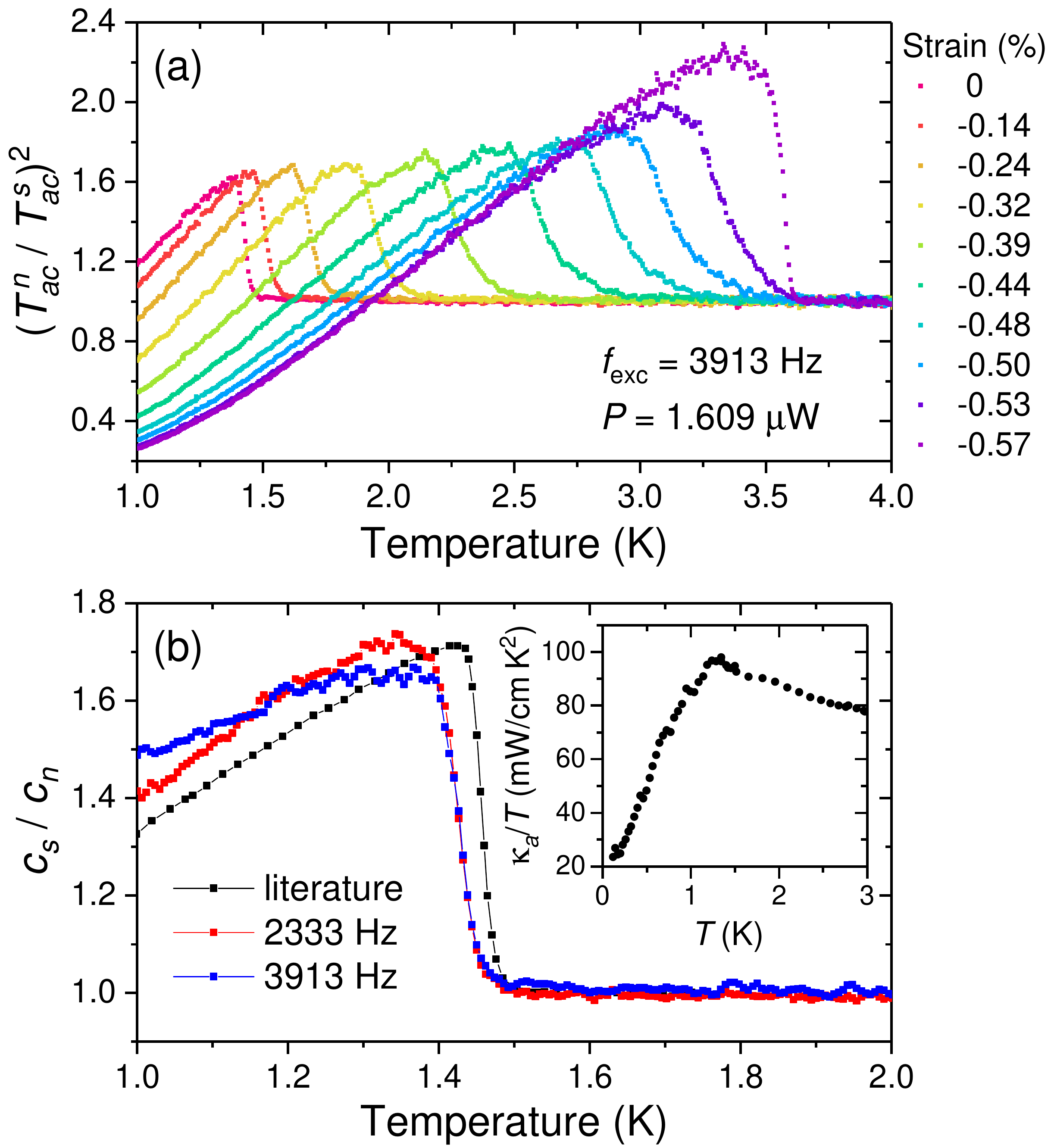}
\centering
\caption{
(a) $(T^{n}_{ac}/T^{s}_{ac})^2$ against temperature for S4 at a series of compressive strains up to the peak in $T_c$. Over a narrow range of temperature across $T_c$ and under the conditions described in the text, $(T^{n}_{ac}/T^{s}_{ac})^2 = c_s/c_n$.
(b) Calculations of normalized specific heat curves at zero strain from our ac calorimetry by inverting Eq.\ \ref{Tac} and using the thermal conductivity from Ref.\ \onlinecite{Hassinger2017} reproduced in the inset. The red and blue curves are the calculations for S4 with $f_{exc} = 2333$~Hz and 3913~Hz, respectively. The black curve is the published data from Deguchi \textit{et al.}\ \cite{Deguchi2004}.
}
\label{CnCs}
\end{figure}

In Figs.\ \ref{HC} -- \ref{HC_smallsteps} we deliberately presented unprocessed data based on our calibrated measurement of the ac temperature change measured by the thermocouple, because those raw data already enable us to prove the existence of bulk superconductivity and search for transitions between different superconducting order parameters.  Going further requires data processing, and in Fig.\ \ref{CnCs} we show two examples of what can be done.

Under an assumption that the heater is narrow and the sample width is small in comparison with the thermal diffusion length, the signal is related to the true volume heat capacity $c_v$ by
\begin{equation}
  T_{ac}=\frac{P\times F(\omega)}{2A\sqrt{\omega}\sqrt{2\kappa(T)c_v(T)}},
\label{Tac}
\end{equation}
in which $A$ is the cross-sectional area of a long bar and $\kappa$ is the thermal conductivity.  In principle, a full temperature dependence of $c_v$ could be calculated by inverting this expression if the thermal conductivities were known at all temperatures and strains.  In practice the strain-dependent thermal conductivities are not known, but the strain evolution of the size of jump in true heat capacity $\Delta c_v$ as a function of strain, normalized to the normal state value, can be tracked by plotting the ratio $(T^{n}_{ac}/T^{s}_{ac})^2$, where $T^{n}_{ac}$ and $T^{s}_{ac}$ are the temperature oscillation amplitudes observed in the normal and superconducting states. At the superconducting transition, $c_v$ has a discontinuous jump, while all other parameters in Eq.\ \ref{Tac} vary continuously. Over a narrow range of temperature we can therefore compare the normal-state and superconducting heat capacities using the relation $(T^{n}_{ac}/T^{s}_{ac})^2 = c_s/c_n$. This is depicted in Fig.\ \ref{CnCs}a.  The data well below the transition contain systematic uncertainty because of the unknown change in $\kappa(T)$ but since $\kappa$ remains continuous at the transition itself, the information about $\Delta c_v / c^n_v$ is reliable.  This ratio is observed to grow slightly as $T_c$ is increased.

Although we do not have sufficient information to invert Eq.\ \ref{Tac} at all strains, enough is known about $\kappa(T)$ at zero strain to allow this to be attempted (see inset of Fig.\ \ref{CnCs}b and Ref.\  \onlinecite{Hassinger2017}).  The results, shown in Fig.\ \ref{CnCs}b, demonstrate the level of accuracy that can be achieved using the present set-up (for which Eq.\ \ref{Tac} is not a perfect description, as discussed in the  Supplemental Material \cite{suppl}).
These results prove that although the full deduction of the volume heat capacity is not necessary for the physics that we sought to investigate in the current experiments, it will be possible in principle to determine it with good accuracy while applying uniaxial stress using the basic technique discussed here.

\section{DISCUSSION}

The work that we have reported in this paper establishes the strain tuning of superconductivity Sr$_2$RuO$_4$ to be a bulk phenomenon, and is also relevant to the ongoing discussion about the nature of its superconducting order parameter.  The three main experimental findings presented above are a) the lack of superconducting transition splitting resolved across the whole measured range of strains studied, which include those crossing the maximum in $T_c$, b) the lack of evidence for any competing transition to other forms of order, and c) the fact that the heat capacity jump size normalized to the value above $T_c$, $\Delta C/C$, grows as Sr$_2$RuO$_4$ is strained toward its maximum $T_c$ of 3.5~K.  We discuss each in turn.

The lack of observed transition splitting invites a discussion of the resolving power of our experiment.  First we examine our ability to resolve a split between two transitions of similar strength that lie close in temperature.  Here the important observation is the lack of structure in the main heat capacity jump as $T_c$ is tuned.  We observe a broadening of the measured anomaly as the strain is increased.  However, the fact that the broadening then becomes less and less pronounced as we approach the maximum in $T_c$ for which $dT_c/d\varepsilon=0$ strongly suggests that it is associated with strain inhomogeneity of $\sim 8$\% (see the Supplemental Material \cite{suppl}).  This leads to transition widths of approximately 50~mK at low strains and 300~mK when $T_c = 3.5$~K.  Transition splitting considerably less than these widths would be hard to resolve; a sensible resolution estimate is approximately 20~mK at low strain and 120~mK at $T_c = 3.5$~K.

The second aspect of the search for a transition in the superconducting state is informed by intriguing recent results from muon spin relaxation measurements under strain.  The key finding of those experiments is that the muon signal that is widely interpreted as evidence for the onset of time reversal symmetry breaking is always observed in a narrow range of temperatures, $T_{\mu}$, around 1.3~K as $T_c$ is increased from 1.5~K to 3.5~K \cite{Grinenko2020}.  A strain-dependent splitting between $T_c$ and $T_{\mu}$ is a key expectation for an order parameter whose time reversal breaking results from two order parameter components summed in the form $p + ip$ \cite{Rice1995}, $d + id$ \cite{Zutic2005,suh2019}, $d +is$ \cite{Romer2019} or $d + ig$ \cite{Kivelson2020}.  The high-resolution, multiply averaged data shown in Fig.\ \ref{HC_highRes} are ideal for careful examination for signs of a thermodynamic signature of $T_{\mu}$.  The noise level achieved there corresponds to sub-picovolt fluctuations of the thermocouple voltage used to deduce the temperature changes at the sample, which we believe that represents the state-of-the-art for measurements of this kind.  As stated in the experimental section, this gives us a detection sensitivity of a few per cent for a thermodynamic transition at $T_{\mu}$ of a similar width to that at $T_c$, \textit{i.e.}\ any second transition larger than a few per cent of the primary one at $T_c$ would be resolvable.  As can be seen from Fig.\ \ref{HC_highRes}, none is resolved in our experiment.  This observation does not rule out the existence of the second transition but it places strong constraints on it.  Arguments are provided in Ref.\ \onlinecite{Grinenko2020} that for order parameters such as $d\pm id$, in which the two components are symmetrically equivalent, a small heat capacity anomaly might be obtained by invoking strong competition between these components, such that condensation of one component suppresses the energy gain when the second condenses. This would require fine-tuning the theory beyond anything required by symmetry. For $d+is$ and $d+ig$ orders, the two components are not symmetrically equivalent, so again, fine tuning is required, this time for them to have similar $T_c$ in unstrained material.  As pointed out in \cite{Kivelson2020}, a $d_{x^2-y^2} + ig$ state in which the $g$ component forms below $T_c$ is qualitatively expected to have a suppressed heat capacity anomaly at this second transition because both components have vertical line nodes along the $[110]$ direction, but whether this is a big enough effect to be consistent with our findings remains unclear. Our data therefore strongly motivate concrete predictions of the size of the transition at $T_{\mu}$ on the basis of any model for time reversal symmetry breaking order parameters in Sr$_2$RuO$_4$.

Similar considerations apply to thermodynamic signatures associated with any proposed magnetic transition if it occurs at lower strains than those required to reach $T_c=3.5$~K.  We do not have the same resolution as that discussed above at all strains and temperatures, but can confidently rule out any transition with a thermodynamic signature of more than 8\% of that of the primary superconducting transition. As $T_c$ drops when the crystal is strained beyond the value required to maximize $T_c$, we confirm previous magnetic susceptibility observations \cite{Steppke2017} that the $T_c$ vs strain curve is strongly asymmetric. $T_c$ drops much more rapidly on the high strain side, as is evident from Figs.\ \ref{HC}b and \ref{HC}c.  This drop is so rapid that our signal inevitably becomes more strongly broadened due to the small strain inhomogeneity discussed above, complicating its interpretation.  However, qualitative inspection of the data in Fig.\ \ref{HC}b indicates that the heat capacity anomaly is shrinking rapidly as well as broadening.  While not conclusive, this would be consistent with the recent report \cite{Grinenko2020} that at very high strains, density wave order sets in above our temperature range of measurement and co-exists with the superconductivity.  Although challenging, further investigation of thermodynamic properties at these strains of over $0.6\%$ will be desirable.

The recent NMR \cite{Pustogow2019,Ishida2019}, polarized neutron scattering \cite{Petsch2020} and upper critical field measurements \cite{Steppke2017} on Sr$_2$RuO$_4$ strongly favor an order parameter with even parity.  This interpretation is consistent with our findings on the growth of $\Delta C/C$ with increasing strain (Fig.\ \ref{CnCs}a). Our observation is significant because it gives qualitative information on the region of the Brillouin zone in which the gap is maximized.  There is strong evidence \cite{Steppke2017,Barber2018,Sunko2019} that the uniaxial-pressure-induced increase in $T_c$ of Sr$_2$RuO$_4$ is associated with tuning its electronic structure to a van Hove singularity.  In these circumstances the density of states near the van Hove point grows, presumably influencing the increase in $T_c$.  However, the superconducting gap does not necessarily have to be maximized in the same region of the Brillouin zone.  Indeed, for any odd parity state, it must be zero by symmetry at the van Hove point, precisely where the density of states will be largest.  Similar considerations would apply to any even parity state with nodes along the $[010]$ direction.  In these circumstances, a strain-dependent drop in $\Delta C/C$ would be qualitatively expected, and the $T_c$ increase would have to be due to a pressure-dependent increase in the pairing potential somewhere else in the Brillouin zone.
In contrast, if the order parameter is allowed to maximize along the $[010]$ direction, $\Delta C/C$  will not drop with increasing strain.  It will either remain constant or rise, because the order parameter efficiently profits from the increased density of states.  The rise in $\Delta C/C$  that we observe therefore favors even parity states without nodes or gap minima along $[010]$ (see for example Ref.~\onlinecite{Sharma2020}), and invites full calculations using concrete models \cite{Zutic2005,Romer2019,Roising2019,Kivelson2020} for states falling into this general class.

Finally, we stress that all of the physics discussed here can be addressed from analysis of the 'raw' data presented in Figs.\ \ref{HC} -- \ref{CnCs}a.  In the long term it will be desirable to incorporate in-situ measurement of the thermal conductivity to enable full quantitative measurement of the heat capacity, but the quantities discussed above do not require this full quantitative analysis.  Their measurement has depended on the very high precision with which our experiment has been performed, and the resultant sensitivity that we have to changes in the heat capacity.

\section{CONCLUSION}

In summary, we have presented high frequency measurements that give us access to the dependence of the heat capacity of Sr$_2$RuO$_4$ under high uniaxial pressures.  The data we have obtained provide important insights into the nature of the superconducting order parameter and provide rigorous thermodynamic constraints which will have to be satisfied by any detailed theoretical proposals for the microscopic nature of the superconductivity.  We believe that the techniques that we have introduced will also be of relevance to thermodynamic studies of other materials under uniaxial pressure.

\begin{acknowledgments}
We thank I.\ I.\ Mazin for useful discussions and R.\ Borth, M.\ Brando and U.\ Stockert for experimental support. NK acknowledges the support from JSPS KAKENHI (nos JP17H06136 and JP18K04715) and JST-Mirai Program (no. JPMJMI18A3) in Japan and YM from JSPS KAKENHI (nos JP15H05852, JP15K21717) and JSPS core-to-core programme.  YSL acknowledges the support of a St Leonard’s scholarship from the University of St Andrews, the Engineering and Physical Sciences Research Council via the Scottish Condensed Matter Centre for Doctoral Training under grant EP/G03673X/1, and the Max Planck Society.
\end{acknowledgments}

\bibliography{Sr2RuO4}

\end{document}